\documentclass[qsecnum,amsmath,preprintnumbers,superscriptaddress,nofootinbib,aps,11pt,a4paper]{revtex4}
\usepackage[utf8]{inputenc}
\usepackage[english]{babel}
\usepackage{amsmath}
\usepackage{amsfonts}
\usepackage{amssymb}
\usepackage{amsthm}
\usepackage{hyperref}
\usepackage{color}
\usepackage{bm}
\usepackage{tikz}

\usetikzlibrary{matrix}

 \def\be{\begin{equation}}
\def\ee{\end{equation}}
 \def\ba{\begin{align}}
\def\ea{\end{align}}
\def\bea{\begin{eqnarray}}
\def\eea{\end{eqnarray}}

\def\a{\alpha}
\def\b{\beta}

\def\m{\mu}
\def\n{\nu}

\begin{document}

\title{{\bf Frame (In)equivalence in Quantum Field Theory and Cosmology }}
\author{ Kevin Falls}
\address{Scuola Internazionale di Studi Superiori Avanzati (SISSA)\\ Via Bonomea 265, 34136 Trieste, Italy.}
\address{INFN, Sezione di Trieste, Italy}
\author{Mario Herrero-Valea}
\address{Institute of Physics, LPPC\\ \'Ecole Polytechnique F\'ed\'erale de Lausanne\\ CH-1015 Lausanne, Switzerland}
\date{\today}

\begin{abstract}
We revisit the question of frame equivalence in Quantum Field Theory in the presence of gravity, a situation of relevance for theories aiming to describe the early Universe dynamics and Inflation in particular. We show that in those cases, the path integral measure must be carefully defined and that the requirement of diffeomorphism invariance forces it to depend non-trivially on the fields. As a consequence, the measure will transform also non-trivially between different frames and it will induce a new finite contribution to the Quantum Effective Action that we name \emph{frame discriminant}. This new contribution must be taken into account in order to asses the dynamics and physical consequences of a given theory. We apply our result to scalar-tensor theories described in the Einstein and Jordan frame, where we find that the frame discriminant can be thought as inducing a scale-invariant regularization scheme in the Jordan frame.
\end{abstract}

\maketitle
\newpage
\tableofcontents
\newpage
\section{Introduction}

A fundamental property that all sensible physical theories share is the fact that physical statements cannot depend 
the choice of variables we use to describe the physical system, even though there maybe be a set of variables which have a preference. For example, in Special Relativity we have the notion of different inertial frames associated to observers moving at different relative velocities. Both observers have their own preferred coordinate frames in which to describe events but physical statements are invariant under Lorentz transformations which relate the two frames. Moving to the theory of General Relativity, we demand physical statements to be invariant under quite arbitrary coordinate transformations on space-time. In classical field theory one can also extend the notion of general covariance to field space by demanding that physical statements are independent of the way we parametrise the field variables. Invariably, the equations of motion will appear simpler if we use a certain set of variables, however the physics should be indifferent to this choice.

Quantum Field Theory (QFT) is a different story though, since the formalism is drastically different to classical mechanics. In perturbative QFT we are interested in amplitudes between asymptotic states, which can be obtained by taking variational derivatives of the Quantum Effective Action after performing a path integral over all possible paths with the right boundary conditions. One problem is that the standard definition of the Quantum Effective Action depends on the choice of variables as a consequence of the source term. However since the source is equal to the effective equations of motion, the non-equivalent pieces which arise for this reason  do not contribute to on-shell amplitudes corresponding to the S-matrix elements.
More generally, since observables are evaluated for vanishing source this dependence on the choice of variables is innocuous. 
Indeed one may even overcome this problem off-shell by using the unique effective action \cite{Vilkovisky:1984st} which makes use of a covariant source term. However, as we shall see, this is not the end of the story. Even with the vanishing source terms, the path integral measure must also transform in a covariant manner for theories formulated with different field variables to be equivalent.
This issue becomes especially subtle in the presence of gravity and whenever extra symmetries are required for the field manifold.

A first indication that the choice of variables can be significant was found in 
 \cite{Herrero-Valea:2016jzz} where it was pointed out that in certain scalar-tensor theories it is possible to map anomalous symmetries (scale invariance) to healthy ones (shift symmetry) after a field redefinition. Specifically, theories which are classically scale invariant in the Jordan frame and are related to a theory which enjoys shift symmetry in the Einstein frame.
  In that case, quantization in the two different sets of variables lead to a different S-matrix due to the appearance of new transition amplitudes only in the Jordan frame, where the scale invariance is anomalous\footnote{The textbook example of the triggering of new S-matrix elements by anomalous currents is the decay of the neutral pion in two photons due to the axial anomaly in chiral perturbation theory.}. In the Einstein frame the shift symmetry remains intact in the quantum theory and consequently no anomaly occurs.  Through the example in \cite{Herrero-Valea:2016jzz}, one can trace the origin of the discrepancy to the existence of the metric as a dynamical degree of freedom, since it is the metric redefinition what eventually leads to the transmutation of symmetries. This prompts us to further investigate the frame dependence of more general scalar-tensor theories and to identify the origin of disparity between different quantum theories.

Although we do not have a complete theory of Quantum Gravity, there are several regimes in which we require the metric to be a dynamical degree of freedom and General Relativity (GR), or alternative theories, to be quantized as an effective field theory. The most prominent of these regimes is inflation, happening at the very first moments of our Universe's lifetime, when the mean energy was high enough for quantum gravitational effects to be of relevance. We do not dispose of an accurate description of inflation though, due to a lack of data to pinpoint a particular theory \cite{Akrami:2018odb}, but instead there exist many models which satisfy the requirements that lead to a successful inflationary regime compatible with our meager data \cite{Starobinsky:1980te,Guth:1980zm, Kallosh:2013hoa, Linde:2015uga, Linde:1983gd,Boubekeur:2005zm,Pajer:2013fsa,Silverstein:2008sg,McAllister:2008hb,Bezrukov:2007ep,Rubio:2018ogq,Bezrukov:2012hx,Fumagalli:2017cdo, Barvinsky:2009ii,Barvinsky:2008ia}. Many of these models are formulated, either explicitly or effectively after disentangling the relevant degrees of freedom, as single field inflation models, where a scalar field is coupled to gravity and moves down a potential, producing inflation while rolling down and stabilizing in the minimum of the potential afterwards. Although this leads to a large zoology of different models, even just for single field inflation, they all share a basic structure on their Lagrangian\footnote{Here and throughout we write the Euclidean  Lagrangians. The corresponding Lorentzian Lagrangian comes with a relative minus sign for each term.}
\begin{align}\label{eq:simple_L}
{\cal L}= -  U(\phi) R+\frac{1}{2}\partial_\mu \phi \partial^\mu \phi + V(\phi)
\end{align}
Different choices of the scalar field potential $V(\phi)$ and the gravitational coupling $U(\phi)$, which includes both the Lagrangian for gravity and the non-minimal interaction terms between gravity and the scalar field $\phi$, will lead to the different explicit proposals for inflation\footnote{There are models that are not explicitly captured by this simple Lagrangian (Starobinsky inflation \cite{Starobinsky:1980te}, for example). However we can get them by minor modifications of \eqref{eq:simple_L}.}. Related to scalar tensor models are $f(R)$ models where ${\cal L} = f(R)$ which, apart for the case where $f(R)$ is linear, also describe one physical scalar particle coupled to a spin-two graviton.

It is of common practice, though, to use field redefinitions to eliminate unpleasant non-minimal couplings in $U(\phi) R$ between the metric and the scalar field e.g. $\phi^2 R$. By redefining a new metric and scalar field, which we denote with tildes, it is always possible to get rid of these terms and arrive to a minimally coupled theory
\begin{align} \label{eq:Einstein_L}
{\cal L}=-M_p^2\tilde{R}+\frac{1}{2}\partial_\mu \tilde{\phi} \partial^\mu \tilde{\phi} + \tilde V(\tilde\phi)
\end{align}
where the gravitational sector is described by plain General Relativity. The theory described by the original Lagrangian \eqref{eq:simple_L} is referred to as the Jordan frame where as the minimally coupled theory is known as the Einstein frame.
Similarly one can also use field redefinitions to rewrite any $f(R)$ model as a scalar tensor theory either in the Jordan frame or the Einstein frame. 
 
The simpler setting of the Einstein frame allows for an also straightforward  interpretation of the dynamics of the system as a scalar field rolling down the new potential, from which we can derive all relevant inflationary parameters. However, as we have previously pointed out, this is a dangerous step if we want to include quantum effects, since the S-matrix of both theories might be different in certain cases and we might be missing important physical effects. Indeed, the question of equivalence of scalar-tensor theories in Cosmology has been thoroughly studied in the recent years from many different points of view (\cite{Capozziello:1996xg,Nojiri:2000ja,Kamenshchik:2014waa,Capozziello:2010sc, Postma:2014vaa,Banerjee:2016lco,Pandey:2016unk,Alvarez:2001qj,Karam:2017zno,Pandey:2016jmv,Bounakis:2017fkv,Karam:2018squ,Faraoni:1999hp} and references therein). However, most works are focused on the classical and observational aspects of frame equivalence and the few that study the issue at a quantum level find contradicting results.  Several works have concentrated on the divergent part of the one-loop the effective action and the corresponding beta functions (for related non-perturbative studies using the functional renormalisation see \cite{Benedetti:2013nya,Ohta:2017trn}) .   These studies show that the divergent part of the effective action generically differs in the two frames by terms proportional to the equations of motion. This was shown in two and four dimensional dilatonic gravity in \cite{Nojiri:2000ja, Alvarez:2014qca} while in four dimensions this has also been proven in \cite{Kamenshchik:2014waa} for a wide range of models.  In \cite{Bounakis:2017fkv} calculations were carried out using the field space-covariant Vilkovisky-DeWitt effective action which guarantees that results are formally independent of parameterisation of the quantum fields. However one should bear in mind that even if one uses a covariant approach results can still depend on the definition of the geometric objects such as metrics and connections defined on field space.   

Motivated by these concerns, which might have consequences for many important inflationary and gravitational models, we wish to revisit the problem of equivalence of Quantum Field Theories. We will do this by giving a proper definition of all the elements involved in the path integral quantization of a given Quantum Field Theory and studying their behaviour under a change of frames. We will find that, as hinted by the previous discussion about anomalies, the source of the apparent inequivalence between the frames is the definition of the path integral measure, which includes the determinant of a metric defined on the field manifold. While this metric is generically field independent for scalars, fermions and vector fields, and thus it can be ignored for perturbative computations, this is no longer the case when gravity enters into the game. The requirement of diffeomorphism invariance of the Quantum Effective Action (even when the metric is just a semi-classical degree of freedom or a external source) forces the integration measure to depend non-trivially on the field variables. Thus, if we want preserve frame equivalence at the quantum level the measure must also transform non-trivially after a change of frames. 
However if we first change frames at the classical level and then quantize the resulting theory the measure will not coincide with the transformed one such that the operations of changing frames and quantizing do not commute. 
Consequently the corresponding Quantum Effective Actions will differ by a non-vanishing finite piece which is not proportional to the equations of motion. This \emph{frame discriminant} term will contribute to 1PI correlation functions and thus it cannot be ignored. Disregarding it represents a different choice of integration measure, and thus a \emph{different Quantum Field Theory}.

This paper is organized as follows. In section \ref{sec:frame_anomaly} we will introduce the concept of frame equivalence both at the classical and quantum level, discussing the state-of-the-art of the discussion and raising some concerns for scalar-tensor theories. In section \ref{sec:functional_integral} we will define the path integral and the integration measure for a general theory, keeping in mind the scalar-tensor theories of interest and discussing the transformation of the path integral measure. 

 We will then present the derivation of the frame discriminant using the background field method in section \ref{sect4} and we will apply our formalism to scalar-tensor theories in section \ref{sec:discr_scalar_ten}, describing also its relation with the so called \emph{scale-invariant regularization}. Finally, we will summarize and discuss our results and conclusions in section \ref{sec:conclusions}. Appendix \ref{gauge_fixing_App} will be devoted to proof some statements about our derivation in the presence of gauge invariance.

\section{Frame equivalence}\label{sec:frame_anomaly} 
Frame equivalence is an important assumption for physics to be reliable. It means that the choice of variables used to describe a system should not matter when deriving physical statements, although of course computations might be simpler for some of these choices than for others. The trivial example of this situation is the case of a particle forced to move in a circumference in classical mechanics. The system can be described either by using Cartesian or polar coordinates. The equations are simpler in the latter but physical statements are equivalent and in one-to-one correspondence, provided that we properly transform quantities between different coordinates systems.

In classical field theories we can give a solid definition of this statement. If we have two frames (two choices of dynamical field variables) $\Phi^a(x)$ and $\tilde{\Phi}^{a}(x) $ related locally by 
\begin{align}\label{eq:equivalence_relation}
\tilde{\Phi}^{a}(x)=\tilde{\Phi}^{a}(\Phi^a(x)),
\end{align}
we call them \emph{equivalent} if any physical quantity $A(\Phi)$ satisfies 
\begin{align}\label{eq:equivalence_condition}
\tilde{A}(\tilde{\Phi})|_{\tilde{\Phi} = \tilde{\Phi}(\Phi)}=A(\Phi).
\end{align}
This is no more than the statement of covariance under the manifold spanned by all possible configurations of the variables $\Phi^a$. Alternatively, we could also say that this defines the notion of what we consider a physical quantity for a general theory. In particular, it encloses a notion of relativity familiar from  General Relativity but where in \eqref{eq:equivalence_condition} the coordinates are the dynamical variables which parametrise the physical system rather than space-time coordinates. We can therefore identify the variables $\Phi$ with coordinates on the space of dynamical histories $\mathcal{M}_\Phi$ such that  \eqref{eq:equivalence_condition} is just the statement of general covariance, where physical observables $A(\Phi)$ are understood as scalars on  $\mathcal{M}_\Phi$.

A very important consequence of what we have described is the equivalence of classical field theory under redefinitions of the variables $\Phi$. This follows from the fact that the action $S(\Phi)$ satisfies \eqref{eq:equivalence_condition}, which means that classical equations of motion, obtained by applying the variational principle to  either $S(\Phi)$   or $\tilde{S}(\tilde{\Phi})$, are related by
\begin{align}\label{eq:deltaS}
\frac{\delta S[\Phi]}{\delta \Phi^a} =    \sum_{b} \frac{\partial \tilde{\Phi}^{b}}{\partial \Phi^a}  \frac{\delta \tilde{S}[\tilde{\Phi}]}{\delta \tilde{\Phi}^{b}}  \,.
\end{align}

Provided that the Jacobian matrix $ \frac{\partial \tilde{\Phi}^b}{\partial \Phi^a} $ is non-singular when evaluated at each of the stationary solutions,  $\frac{\delta S(\Phi)}{\delta \Phi^a}=0$ implies $ \frac{\delta \tilde{S}(\tilde{\Phi})}{\delta \tilde{\Phi}^b}=0$ and stationary trajectories are in a one-to-one correspondence. One can then say that two theories where the variables are related to each other by \eqref{eq:equivalence_relation} are classically equivalent if the dynamical shells corresponding to the points $\Phi_0$ and $\tilde{\Phi}_0$, where the actions are stationary, are related by $\tilde{\Phi}_0 = \tilde{\Phi}(\Phi_0)$. Thus, any physical quantity will lead to the same result in either frame when evaluated on-shell, as a consequence of \eqref{eq:equivalence_condition}.
Again, this just encloses the common notion that one should be free to choose whatever variables they prefer to perform a computation and, although particular equations will be different, physical statements must remain the same for any choice.

Although this statement is crystal clear in classical mechanics, the situation is not so transparent in Quantum Field Theory(QFT), where not only stationary trajectories contribute to the dynamics of a given system. There, instead, we are interested in objects formally obtained from a path integral over all possible trajectories with the right boundary conditions. In particular, we focus our interest on correlation functions obtained from the Quantum Effective Action $\Gamma[{\cal Q}]$, which is defined in terms of the mean field ${\cal Q}$ by a Legendre transform
\begin{align}
\Gamma[{\cal Q}]=W[\mathcal{J}]-\mathcal{J}\cdot{\cal Q}
\end{align}
where ${\cal Q}$ satisfies the effective equations of motion
\begin{align}
\frac{\partial \Gamma}{\partial {\cal Q}^a}=-\mathcal{J}_a
\end{align}
and the effective potential is given as a (Euclidean) path integral over the field variables with a source $\mathcal{J}_a$
\begin{align}
{\cal Z}[\mathcal{J}]=e^{- W[\mathcal{J}]}=\int [d\Phi] \,e^{-S[\Phi]- \mathcal{J}\cdot \Phi}
\end{align}

Here the dot product is assumed to represent sum over all indices as well as integration over space-time coordinates
\begin{align} \label{Source_term}
\mathcal{J}\cdot \Phi =\int d^4 x \, \mathcal{J}_a \Phi^a \,.
\end{align}

However, as noted by Vilkovisky in \cite{Vilkovisky:1984st}, the quantum effective action as defined here does not satisfy an analogous formula to \eqref{eq:deltaS}. In general
\begin{align} \label{Non-covariance_Gamma}
\frac{\delta \Gamma[{\cal Q}]}{\delta {\cal Q}^a} \neq \sum_b \frac{\delta \tilde{{\cal Q}}^b}{\delta{\cal Q}^a}  \frac{\delta \tilde{\Gamma}(\tilde{{\cal Q}})}{\delta \tilde{{\cal Q}}^b}  
\end{align}
and there is not a one-to-one correspondence of 1PI correlation functions in different frames. Nevertheless, it was also shown in \cite{Vilkovisky:1984st} that the problematic pieces are proportional to the equations of motion and they cancel on-shell, preserving equivalence for those correlators that contribute to S-matrix elements. The problem persists off-shell, and although there is a way to covariantize $\Gamma[{\cal Q}]$, arriving to what is known as Unique Effective Action, it is not clear if this redefinition is needed at all for most standard settings, since all dynamics is presumably contained in the S-matrix\footnote{There is some discussion about the need of using the Unique Effective Action in order to obtain gauge invariant beta functions for running couplings \cite{Barvinsky:1984jd, Jacopo}.}. One may summerize the situation by noting that it is the non-covariance of \eqref{Source_term} which is responsible for \eqref{Non-covariance_Gamma} but that, since observables are calculated for ${\cal J} = 0$, this can only lead to disparities in intermediate steps in the calculation of observables (e.g. correlation functions) but not in the final result (e.g. the S-matrix).

Even though things seem pretty clear from Vilkovisky's arguments, more concerns can be raised in the presence of gravity as one of the dynamical fields in $\Phi$, even for on-shell quantities. In particular, let us focus in the problem pointed out in \cite{Herrero-Valea:2016jzz}, where we consider non-linear redefinitions of the fields. In those cases, the realization of gauge and global symmetries might differ in different frames. Therefore, it might also happen that something which is an exact symmetry under renormalization in one frame maps to an anomalous symmetry in the other. Then, the anomaly to the current conservation generates new S-matrix elements in one of the frames only, through the expectation value
\begin{align}
\langle 0 |\nabla_\mu J^\mu|0\rangle=\langle 0|\Psi\rangle \neq 0 
\end{align}
where $|\Psi\rangle=\nabla_\mu J^\mu|0\rangle$ is some state of the theory. This amplitude, which it is not generated in the second frame, where there is no anomaly, spoils the equivalence premise in a strong way.

Although this is a quite general effect associated to field redefinitions, let us here be explicit and show a realization of this phenomenon by choosing a particular scale-invariant scalar-tensor theory, where we couple the metric $g_{\m\n}$ to a scalar field $\phi$ in the Jordan frame
\begin{align}\label{eq:jordan_frame}
S_J[g_{\m\n},\phi]=\int d^4 x \sqrt{|g|}\left(-\xi \phi^2 R+\frac{1}{2}\partial_\mu \phi \partial^\m \phi +\frac{\lambda}{4!}\phi^4   \right)
\end{align}
where $\xi$ and $\lambda$ are dimensionless couplings. This action is invariant under diffeomorphisms as well as under global scale transformations of the form
\begin{align}\label{eq:scale_inv}
g_{\m\n}\rightarrow \Omega^2 g_{\m\n} ,\qquad \phi \rightarrow \Omega^{-1}\phi
\end{align}
for constant $\Omega$. This symmetry is extended to local Weyl invariance when $\xi=-\frac{1}{12}$, for which the scalar field becomes a gauge degree of freedom \cite{Alvarez:2014qca}. This defines our first frame, with variables collectively denoted as $\Phi$. 

The frame $\tilde{\Phi}$ will be the corresponding Einstein frame, obtained by defining a new of set of variables through
\begin{align}\label{eq:transf}
\tilde{g}_{\m\n}=\frac{\xi \phi^2}{M_p^2}g_{\m\n}, \qquad \tilde{\phi}=M_p\sqrt{\frac{1}{\xi} +12}\, \log\left(\frac{\phi}{m}\right)
\end{align}
where we have introduced two new scales, the Planck mass $M_p$ and an arbitrary scale $m$. The corresponding action is
\begin{align}\label{eq:Einstein_frame}
S_E[\tilde{g}_{\m\n},\tilde{\phi}]=\int d^4 x \sqrt{|\tilde{g}|}\left(- M_p^2 \tilde{R} +\frac{1}{2}\partial_\mu \tilde{\phi}\partial^\mu\tilde{\phi} +\frac{\lambda}{4!}\frac{ M_p^4}{ \xi^2}\right)
\end{align}

In this frame, the action is invariant under diffeomorphisms too but, instead of enjoying a scale symmetry, this has mutated into a shift symmetry for the scalar field
\begin{align} \label{eq:Shift}
\tilde{\phi}\rightarrow\tilde{\phi}+ C
\end{align}
where $C$ is a constant. 

Now we can ask to what extent the two actions \eqref{eq:jordan_frame} and \eqref{eq:Einstein_frame}  are classically equivalent.
If we consider the equations of motion for \eqref{eq:jordan_frame} it is clear that $\phi =0$ is a solution for all metrics $g_{\mu\nu}$.
However for $\phi = 0$ the coordinate transformation between the two frames is singular, since it maps to $\tilde{g}_{\mu\nu} = 0$ and $\tilde{\phi} = \infty$. Thus, equivalence demands the theory to be in the broken phase. As long as we give a vacuum expectation value to the field $\phi$, both frames are classically equivalent.

However, this is not the end of the story. As we commented, this setup gives us an explicit relalization of the problem referred to as anomalous frame equivalence in \cite{Herrero-Valea:2016jzz}. When quantizing this theory in the Jordan frame in dimensional regularization\footnote{Here we use dimensional regularization for simplicity of the discussion and computations, since it is a standard tool in QFT. However, any other regularization will unavoidably lead to the same conclusions.} we will generate contributions to the effective action of the generic form
\begin{align}
\Gamma_{J}[g_{\m\n},\phi]=\frac{1}{\epsilon}\int d^dx \sqrt{|g|}\sum {\cal O}_J[R_{\m\n\a\b},\phi]+\text{finite}
\end{align}
where ${\cal O}_J[R_{\m\n\a\b},\phi]$ are homogeneous operators of energy dimension $4$, $R_{\m\n\a\b}$ is the Riemann tensor constructed with $g_{\m\n}$ and we have restricted ourselves to a single loop in the perturbative expansion. Note that the volume integral here is $d$-dimensional, with $d=4+\epsilon$. This means that, after a scale transformation, the integrand will transform as
\begin{align}
\sqrt{|g|}\sum {\cal O}_J[R_{\m\n\a\b},\phi] \rightarrow \sqrt{|g|}\sum {\cal O}_J[R_{\m\n\a\b},\phi] \Omega^{\epsilon}
\end{align}
which, after expanding in $\epsilon$ will generate a finite residue on the transformation of the effective action
\begin{align}
\delta \Gamma_{J}[g_{\m\n},\phi]=\omega \int d^dx \sqrt{|g|}\, \sum {\cal O}_J[R_{\m\n\a\b},\phi]
\end{align}
where $\Omega=1+\omega+O(\omega^2)$.

This is the usual scale anomaly of theories with scale invariance in curved space, where new contributions to the gravitational lagrangian are generated by radiative corrections and, whenever we can define a sensible S-matrix\footnote{The definition of the S-matrix depends on the uniqueness of asymptotic states, which is only possible if the space-time is globally hyperbolic \cite{Friedman:1992je}.}, they will generate a new scattering amplitude from the anomalous contribution to the current conservation
\begin{align}\label{eq:current_jordan}
\langle 0 | \nabla_\mu J_J^{\mu}|0\rangle = \sum {\cal O}_J[R_{\m\n\a\b},\phi]
\end{align}
where $J_J^\mu$ is the classically conserved current associated to the symmetry \eqref{eq:scale_inv}.

In the Einstein frame, we can also proceed with quantization in the standard fashion, also using dimensional regularization. In that case, the one-loop effective action will take a similar form
\begin{align}
\Gamma_{E}[\tilde{g}_{\m\n},\tilde{\phi}]=\frac{1}{\epsilon}\int d^dx \sqrt{|\tilde{g}|}\sum {\cal O}_E[\tilde{R}_{\m\n\a\b},\tilde{\phi}]+\text{finite}
\end{align}

The operators ${\cal O}_E$ can be obtained, when on-shell, as a transformation of the corresponding ones in the Jordan frame, satisfying equivalence in the sense of \cite{Vilkovisky:1984st}. However, in this case there is no anomaly in the shift symmetry, so there is no obstruction to the conservation of the corresponding current
\begin{align}
\langle 0 | \nabla_\mu J_E^{\mu}|0\rangle = 0
\end{align}
and the new elements are \emph{not generated}. This effect distinguishes the frames.

There is an obvious clash here with the conclusion of \cite{Vilkovisky:1984st}, which claims that the S-matrix must be equivalent in both frames. However, by examining this particular illuminating example of a scale-invariant theory that we have chosen, it is not difficult to see what is the origin of the issue. Going carefully over the derivation described in the previous paragraphs, we see that there are two crucial steps involved in the computation -- the introduction of a regularization and the computation of the anomaly. Taking into account that, provided that the transformation between frames is not singular, the action transforms in a proper manner, this clearly isolates the origin of the problem in the measure of the path integral. Indeed, if one goes over the derivation of the Unique Effective Action in \cite{Vilkovisky:1984st}, it can be seen that although the path integral measure is carefully defined in the paper, it is considered to be the same in any frame and thus to give the same contribution regardless of the choice of variables. We will see in the next section that this is actually not true.

\section{The functional integral}
\label{sec:functional_integral}
Let us consider the Euclidean path integral of a quantum field theory described by a set of fields collectively denoted as $\Phi$ with Euclidean action $S[\Phi]$. That is
\begin{align}
{\cal Z}[\mathcal{J}]=\int [d\Phi]\,e^{-S[\Phi]-\mathcal{J}\cdot \Phi}\,.
\end{align}
Since the source term $\mathcal{J}\cdot \Phi$ breaks reparameterisation invariance we will from now on consider the case $\mathcal{J} = 0$.
This allows us to concentrate on effects which come from different choices of the functional measure.

Here $[d\Phi]$ represents the functional measure \emph{before regularisation}, whose definition might vary depending on the parametrization of the degrees of freedom used to construct the perturbative expansion of the path integral. However, we must require this measure to be reparametrisation invariant. This can be achieved by regarding the fields $\Phi$ as coordinates on the configuration space $\mathcal{M}_\Phi$. General covariance in this space thus defines 
\begin{align}
[d\Phi]=\prod_a \frac{d\Phi^a}{\sqrt{2\pi}}\sqrt{\det C_{ab}[\Phi]}  V_{\text{gauge}}^{-1}\,.
\end{align}
Let us explain the different objects that appear in this formula. First, we have allowed the action to be invariant under certain symmetry (global or gauge) such that the field $\Phi^a$ is a section of the corresponding bundle, carrying an index which might be also used as a label for the different field species. With the dependence of the field suppressed $\Phi^a = \Phi^a(x)$, we can understand $a$ as a DeWitt index. Consequently the product over $a$ also implies a product over points in spacetime. The factor of $\sqrt{2\pi}$ appears for normalization purposes. The measure is then parametrized by the metric $C_{ab}$ in $\mathcal{M}_\Phi$, which will be generally curved and is to be understood as a two point function of the ultra-local form
\bea \label{C_form}
C_{ab} = C_{ab}(x) \delta(x,y)
\eea
where $C_{ab}(x)$ are local functions of the fields $\Phi^a(x)$.

For gauge theories we also have to divide by the volume of the gauge group of the action. The definition of $V_{\text{gauge}}$ also requires a metric such that
\bea
V_{\text{gauge}} = \int \prod_\a \frac{d\xi^\a}{\sqrt{2\pi}} \sqrt{ \det \eta_{\a\b}[\Phi]}
\eea
where $\xi^\alpha$ are the generators of the Lie algebra of the symmetry. Here again $\alpha$ is understood as a DeWitt index including both the discrete index and the spacetime coordinate and $ \eta_{\a\b}$ has the ultra local form $ \eta_{\a\b} =  \eta_{\a\b}(x) \delta(x,y)$.

Thus the path integral will not depend only on the action $S[\Phi]$ but also on the choice for the metrics $C_{ab}[\Phi]$ and $\eta_{\a\b}[\Phi]$. For theories such as Yang-Mills, the metrics can be chosen to be independent of the fields without breaking gauge invariance and thus they are not relevant for the computation of correlators in perturbation theory. However, in the presence of gravity, the metrics have to depend on the dynamical fields $\Phi$ themselves to preserve diffeomorphism invariance \cite{Fujikawa:1983im}. This implies that in a general case we cannot neglect the contribution of the functional measure into the result of the path integral.

As it is written, the integration measure is reparameterisation invariant, since it is invariant under diffeomorphisms on $\mathcal{M}_\Phi$. Therefore at a purely mathematical level we are free to choose a different parametrisation of our quantum fields where
\bea \label{Phi_tilde}
\tilde{\Phi}^a = \tilde{\Phi}^{a}(\Phi^b)
\eea
Throughout this paper we will be interested in the case where $\tilde{\Phi}^{a}(\Phi^b)$ is a local invertible function of the fields that does not involve derivatives such as \eqref{eq:transf}. 
 Under this change of variables, the action $S$ satisfies \eqref{eq:equivalence_condition} in a trivial manner
\bea \label{Equiv_actions}
\tilde{S}[\tilde{\Phi}] = S[\Phi]
\eea
while the metrics $C_{ab}$ and $\eta_{\a\b}$ transform as a tensor and a set of scalars respectively,
\bea \label{Equiv_metrics}
\tilde{C}_{ab}[\tilde{\Phi}] =  \frac{\delta \Phi^c}{\delta \tilde{\Phi}^a} C_{cd} \frac{\delta \Phi^d}{\delta \tilde{\Phi}^b}\,,  \,\,\,\,\,\,\,   \tilde{\eta}_{\a\b}(\tilde{\Phi})   = \eta_{\a\b}(\Phi) 
\eea
which maintains the form \eqref{C_form} provided that $\tilde{\Phi}^{a}(\Phi^b)$ is a local function of the fields.

We are then free to equivalently write the path integral in the form
\begin{align}
{\cal Z}=\int [d\tilde{\Phi}]e^{-\tilde{S}[\tilde{\Phi}]}\,,
\end{align}
where now
\bea
[d\tilde{\Phi}] = \prod_a \frac{d\tilde{\Phi}^a}{\sqrt{2\pi}}\sqrt{\det \tilde{C}_{ab}[\tilde{\Phi}]}    V_{\text{gauge}}^{-1} \,,
\eea
provided both \eqref{Equiv_actions} and \eqref{Equiv_metrics} hold.
Note that here we could have written $\tilde{{\cal Z}}$ on the left hand side of the previous formula. However, we want to stress the fact that the value of the path integral in the new variables must remain the same, we are just performing a change of variables. Thus, as long as one properly transforms the integration measure, the choice of field variables \emph{can not affect the physics}; it is just a choice of coordinates on $\mathcal{M}_\Phi$. However, what can affect physics is the choice of the metric $C_{ab}$ and the choice of the metric $\eta_{\a\b}$. If one were to choose different metrics $C_{ab}$ and $\eta_{\a\b}$ then evidently the path integral would be different.
Thus, while classically we require that equivalent theories have actions related by \eqref{Equiv_actions}, quantum mechanically we have the addition requirement that the measures of the theories are equivalent which is satisfied by \eqref{Equiv_metrics}. 

The explicit construction of the metric is a subtle issue and different approaches can be found in the literature \cite{Fradkin:1974df,Fujikawa:1983im,Toms:1986sh,Fradkin:1976xa,Falls:2017cze}. A fundamental restriction on the choice of $C_{ab}$ and $\eta_{\a\b}$ which we can impose is that they must lead to a BRST invariant measure for gauge theories \cite{Fujikawa:1983im}. However this only dictates that they transform in a covariant manner under a gauge transformation and does not fix their form. Thus to completely fix the measure we must give a prescription which may itself depend on a preferred choice for the field variables (also phrased as a choice of \emph{frame}). Since different prescriptions lead to different path integrals involving the same action, they correspond to {\it different quantisations} of the same classical theory. In other words we may encounter a situation where \eqref{Equiv_actions} holds but $\eqref{Equiv_metrics}$ is violated.

A prescription which is usually used to determine the metrics, either explicitly or implicitly, is to choose them to cancel ultra-local divergences which appear in the one-loop expression for the path integral. To see how this arises naturally, let us consider the simple example of a free scalar field in curved space-time with action
\bea
S_{\rm free}[\phi,g_{\mu\nu}] = \frac{1}{2} \int d^4x \sqrt{|g|}\,  g^{\mu\nu} \partial_\mu \phi  \partial_\nu \phi
\eea 
We can then write the source-free path integral where we integrate over the fields $\phi$ as
\bea
\mathcal{Z}_{\rm free}[g_{\mu\nu}] = \int d \phi \sqrt{\det C[\phi,g_{\mu\nu} ]} e^{-S_{\rm free}[\phi,g_{\mu\nu}]} \,
\eea
where $d \phi$ is shorthand for $\prod_x \frac{d\phi(x)}{\sqrt{ 2 \pi}}$.

The question then is by what criteria should we fix  $C[\phi,g_{\mu\nu} ]$? First, let us impose that $\mathcal{Z}_{\rm free}[g_{\mu\nu}]$ is diffeomorphism invariant. To ensure that this is the case we can impose that the line element 
\bea
d\ell^2 =  \iint d^4 x\, d^4 y\, \delta\phi(x) C(x,y) \delta \phi(y)  
\eea
is itself diffeomorphism invariant. This implies that $\int d \phi \sqrt{\det C[\phi,g_{\mu\nu} ]}$ is diffeomorphism invariant as well, which along with the diffeomorphism invariance of $S_{\rm free}[\phi,g_{\mu\nu}]$ in turn implies the invariance of $\mathcal{Z}_{\rm free}[g_{\mu\nu}]$. Furthermore if we impose that  $C(x,y)$ is ultra-local, we can determine it up to the choice of a scalar $s(x)$ where
\bea
C(x,y) = \sqrt{|g|}  s(x)  \delta(x,y)
\eea
such that 
\begin{align}
d\ell^2 = \int d^4x\,  s(x) \sqrt{|g|} (\delta\phi(x))^2 \,.
\end{align}

If we assume that $s(x)$ is independent of $\phi$ we can thus formally perform the functional integral to obtain
\bea
\mathcal{Z}_{\rm free}[g_{\mu\nu}] =   \left[\det\left( C^{-1} S^{(2)}\right)\right]^{-\frac{1}{2}} =  \left[\det \left(- s^{-1}(x) \nabla^2\right)\right]^{-\frac{1}{2}}
\eea
where $S^{(2)}$ refers to the Hessian of the action. Then, the natural choice is to take $s(x)= \Lambda^{2}$ to be a positive constant where $\Lambda$ should have the dimension of a mass to ensure that $\mathcal{Z}_{\rm free}$ is dimensionless. The prescription can be thus summarised (and generalised straightforwardly) as identifying  the metric $C_{ab}$ with the coefficient of the Laplacian appearing in the Hessian of the action multiplied by the constant $\Lambda^2$. That is, if we assume that the term in the Hessian involving two derivatives is of the form
\bea
S_{ab}^{(2)} =   - G_{ab} \mathfrak{g}^{\mu\nu} \partial_{\mu} \partial_{\nu} +  \mathcal{O}(\partial)   =- G_{ab} \nabla^2 + ...
\eea 
for some metric $\mathfrak{g}_{\mu\nu}$, we can then choose $C_{ab} =  \Lambda^2 G_{ab}$. We shall refer to this method of determining the measure as the {\it standard} procedure as it is the one which is adopted in practice. A derivation of this prescription starting from the phase space path integral which defines the canonical theory is given in \cite{Toms:1986sh}.

However  there is an ambiguity once we include the space-time metric as one of the quantum fields if there is not a unique choice of which metric we use to construct the Laplace operator $\mathfrak{g}^{\mu\nu} \partial_{\mu} \partial_{\nu} $ in the previous formula. Different choices of metrics will lead to different Laplace operators and to different choices for $G_{ab}$. For instance, different conformally related metrics
\begin{align} \label{conformal_preffered_metrics} 
\mathfrak{g}_{\mu\nu}(\sigma)=e^{2\sigma} \mathfrak{g}_{\mu\nu}(0)
\end{align}
will lead to different definitions of $G_{ab}$ depending on the value of $\sigma$
\begin{align}
G_{ab}(\sigma)=e^{2 \sigma} G_{ab}(0)
\end{align}
If $\sigma$ were field dependent, then we would find that the integration measure depends non-trivially on which metric is identified with $\mathfrak{g}_{\mu\nu}$. 

Thus, the prescription is not unique. Different choices of the preferred space-time metric $\mathfrak{g}_{\mu\nu}$ will lead to different path integrals. This choice can then be interpreted as a preferred frame choice since, if we consider two parametrisations of the fields $\Phi^a$ and $\tilde{\Phi}^a$ which include metrics $g_{\mu\nu}$ and $\tilde{g}_{\mu\nu}$ respectively, then the choices $ \mathfrak{g}_{\mu\nu} = g_{\mu\nu}$ and $\mathfrak{g}_{\mu\nu} = \tilde{g}_{\mu\nu}$ will in general lead to different path integrals. Nonetheless the choice of which field variables we use to carry out the calculation is independent of how we identify $\mathfrak{g}_{\mu\nu}$ in order to determine the form of the measure. While the former choice does not affect the physics, the latter choice can be understood as a different quantization which can lead to different physical predictions and thus to different quantum field theories. One can therefore trace the consequences of defining theories with different preferred metrics $\mathfrak{g}_{\mu\nu}$ to the additional factor of $e^{2 \sigma}$.  By carefully keeping track of this difference one can then identify a concrete physical difference between the two inequivalent quantum theories.

\section{The frame discriminant: A background field approach}\label{sect4}

In the previous section we argued that the choice of variables for the path integral influences the choice of the integration measure, and that in the presence of gravity this can lead to inequivalent contributions to the path integral. We did this in a schematic way, using a simple free theory as a toy example. The purpose of this section is to generalize this result and present a derivation of this effect at one-loop in the perturbative expansion by using the background field method \cite{Abbott:1981ke}. Specifically, we will always have in mind the example of a scalar-tensor theory of the general form
\bea \label{scalar_tensor_action}
S[\phi,g_{\mu\nu}]= \int d^4x \sqrt{|g|}\left[ \frac{1}{2} Z(\phi) \nabla_\mu \phi \nabla^{\mu} \phi   - U(\phi) R  + V(\phi) \right]
\eea
where the fields  $\Phi = \{ g_{\mu\nu}, \phi \}$ are the metric $g_{\mu\nu}$ and a scalar field $\phi$, and of which \eqref{eq:jordan_frame} is a particular example.
Many of the details of the one-loop path integral for this model have been worked out in \cite{Steinwachs:2011zs}.

Here we consider a frame $\tilde{\Phi}^a$ where the fields $\tilde{\Phi} = \{ \tilde{g}_{\mu\nu}, \tilde{\phi} \}$ are related to the original frame $\Phi$ by \eqref{eq:equivalence_relation} which we take to be invertible such that we also have functions  $\Phi^a = \Phi^a(\tilde{\Phi})$.
The action $S$ therefore transforms as a scalar in the sense that
\bea\label{eq:action_related}
\tilde{S}[\tilde{\Phi}] = S[\Phi]|_{\Phi= \Phi[\tilde{\Phi}] }
\eea
where $S$ and $\tilde{S}$ are the actions before gauge fixing. In particular, we will consider that the spacetime metrics will differ by a non-trivial conformal factor 
\bea \label{Conformal_metrics}
\tilde{g}_{\mu\nu} = e^{2 \sigma} g_{\mu\nu}|_{\sigma = \sigma(\phi)}
\eea
for some function $\sigma(x)$ of the space-time coordinates which can be expressed as a function of $\phi(x)$. Again, note that \eqref{eq:transf} is a particular example of this.

Now we note that up to terms proportional to the equations of motion
\bea \label{Hessians_relation}
\frac{\delta^2 \tilde{S}[\tilde{\Phi}]}{ \delta \tilde{\Phi}^{a} \delta \tilde{\Phi}^{b}}     
=     
\frac{\delta \Phi^b}{\delta \tilde{\Phi}^{b}}    \frac{\delta  \Phi^a }{ \delta \tilde{\Phi}^{a}   }  \frac{\delta S[\Phi]}{\delta \Phi^a  \delta \Phi^b}   + \mathcal{O}\left(\frac{\delta S}{\delta \Phi} \right)
\eea
and thus the on-shell Hessian transforms as a tensor on $\Phi$. This is true of the Hessian without the gauge fixing terms, however as we demonstrate in appendix~\ref{gauge_fixing_App},   the relation \eqref{Hessians_relation} remains true when we use the minimal gauges in both of the respective frames. The Hessians of the gauge fixed action have the form
\bea \label{Minimal_Hessian}
 \frac{\delta^2 (S[\Phi] + S_{\rm gf }[\Phi])  }{\delta \Phi^a  \delta \Phi^b}     \equiv     \mathcal{D}_{ab} =   - G_{ab}  g^{\mu\nu} \nabla_{\mu} \nabla_{\nu}  + 2\Gamma_{ab}^\mu \nabla_{\mu} + W_{ab}
\eea
and
\bea
\frac{\delta^2 (\tilde{S}[\tilde{\Phi}]+ \tilde{S}_{\rm gf }[\tilde{\Phi}])}{ \delta \tilde{\Phi}^{a} \delta \tilde{\Phi}^{b}}    \equiv     \tilde{\mathcal{D}}_{ab}  =      - \tilde{G}_{a b}  \tilde{g}^{\mu\nu} \tilde{\nabla}_{\mu} \tilde{\nabla}_{\nu}  + 2\tilde{\Gamma}_{a b}^\mu \tilde{\nabla}_{\mu} + \tilde{W}_{a b} \,.
\eea

Explicitly, the components of $G_{ab}$ in the case of the scalar-tensor theories \eqref{scalar_tensor_action} are given by
\bea
G_{ab} = \begin{pmatrix}
    - \frac{1}{4} U  g^{\mu\nu \, \rho \lambda}  &&  +  \frac{1}{2} U' g^{\mu\nu}     \\   
+\frac{1}{2} U' g^{\rho\lambda} && Z - \frac{(U')^2}{U}    \\
   \end{pmatrix}   \sqrt{|g|}  \delta(x,y) \,.
\eea
 where $ g^{\mu\nu \, \rho \lambda} = g^{\mu \rho} g^{\nu\lambda} + g^{\mu \lambda} g^{\nu\rho} - g^{\mu\nu} g^{\rho \lambda}$.

Now we can ask how $\tilde{G}_{ab}$ will be related to $G_{ab}$. From \eqref{Conformal_metrics} and \eqref{Hessians_relation} it follows that
\bea \label{tildeC_to_C}
\tilde{G}_{a b} =  e^{2\sigma}   \frac{\delta  \Phi^a }{ \delta \tilde{\Phi}^{a}   }   G_{ab}     \frac{\delta  \Phi^b }{ \delta \tilde{\Phi}^{b}   }
\eea
which shows that $\tilde{G}$ and $G$ are inequivalent metrics. Specifically, they differ by the factor $e^{2\sigma}$ in addition to the expected tensor transformation between frames.

As we discussed in the previous section, there now comes a choice of which spacetime metric $g_{\mu\nu}$ or $\tilde{g}_{\mu\nu}$ we select as the physical one $\mathfrak{g}_{\mu\nu}$, since different choices will lead to inequivalent path integral measures.
 If we choose $\mathfrak{g}_{\mu\nu} = g_{\mu\nu}$ the metric on field space is given by $C_{ab} =  \Lambda^2 G_{ab}$ which is the natural measure in the $\Phi$ frame.  Alternatively, if we declare the physical spacetime metric to be $\mathfrak{g}_{\mu\nu} = \tilde{g}_{\mu\nu}$, which is the natural choice in the $\tilde{\Phi}$ frame, the field space metric is given by $\tilde{C}_{ab} =  \Lambda^2  \tilde{G}_{ab}$. However from \eqref{tildeC_to_C} we see that $C_{ab}$ and $\tilde{C}_{ab}$ are not related by simply a change of coordinates on field space.

 The next step is to construct the path integral in both frames. In the $\Phi$ frame the path integral in the minimal background field gauge is given by 
\bea \label{Gauge_fixed_Z}
\mathcal{Z} =   \int d \bar{c} \sqrt{  \det Y_{\a\b} }    \int d c \frac{1}{ \sqrt{ \det \eta_{\a\b}[\Phi]} } \int d \Phi \sqrt{\det C_{ab}[\Phi]} \, e^{-S[\Phi]   - S_{\rm gf}[\Phi]    -\int  d^4x F \bar{c}_{\a} Q^{\a}\,_\b c^{\b} } \,
\eea
where the gauge fixing action is given by $S_{\rm gf} = \frac{1}{2} \int d^4x F^{\a}Y_{\a\b} F^{\b}$. After expanding around a background solution $\Phi_B$, this can be computed at one loop to be
\bea \label{gauge_fixed_Z_result}
\mathcal{Z}  =  e^{-S[\Phi_B]}  \frac{1}{ \sqrt{\det \left[ (C^{-1})^{ac} \mathcal{D}_{cb} \right]}} 
(\det [ Q^{\a}\,_\b])  
\sqrt{\det [(\eta^{-1})^{\a \gamma} Y_{\gamma \b}]}
\eea
One can also construct the minimal gauge in the $\tilde{\Phi}$-frame leading to an analogous expression for the path integral $\tilde{\mathcal{Z}}$.
As we show in appendix~\ref{gauge_fixing_App}, the Fadeev-Popov operators in the two frames and in their respective minimal gauges are also related by 
 \bea \label{FP_relation}
\tilde{Q}^{\a}\,_\b  = e^{-2 \sigma}  Q^{\gamma}\,_\b \,,
\eea
while the $Y$ and $\tilde{Y}$ are related by 
\bea \label{Y_relation}
\tilde{Y}_{\a\b} = e^{4 \sigma} Y_{\a\b}
\eea 
By the choice $\eta_{\a\b} = \Lambda^4 Y_{\a\b}$, the last ultra-local factor in \eqref{gauge_fixed_Z_result} is unity up to factors of $\Lambda^4$ which are needed to ensure that the path integral is dimensionless. However canceling the analogous ultra-local factor in the $\tilde{\Phi}$ frame means that we choose $\tilde{\eta}_{\a\b} = \Lambda^4 \tilde{Y}_{\a\b}$. Thus $\eta_{\a\b}$ and $\tilde{\eta}_{\a\b}$ will also differ depending on which frame the theory is quantized in.

 Defining $ \Delta^{c}\,_b$ by $\mathcal{D}_{ab}  = \Lambda^{-2} C_{ac} \Delta^{c}\,_b$   the path integral is then given by
\bea
\mathcal{Z}  =  e^{-S[\Phi_B]}  \frac{1}{ \sqrt{\det \left[ \Lambda^{-2} \Delta^a\,_b \right]}} 
(\det [  \Lambda^{-2} Q^{\a}\,_\b])  
\eea
where the factors of $\Lambda^4$ appearing in $\eta_{\a\b}$ are used to make the Fadeev-Popov determinant dimensionless. 
Similarly the one-loop path integral in the $\tilde{\Phi}$ frame is given by 
\bea
\tilde{\mathcal{Z}}  &=&  e^{-\tilde{S}[\tilde{\Phi}_B]}  
\frac{1}{ \sqrt{\det \left[ \Lambda^{-2} \tilde{\Delta}^{a}\,_{b} \right]}}
(\det [  \Lambda^{-2} \tilde{Q}^{\a}\,_\b])\\
&  =&    e^{-\tilde{S}[\tilde{\Phi}_B]}  
\frac{1}{ \sqrt{\det \left[ \Lambda^{-2} e^{-2 \sigma}  \Delta^{a}\,_{b} \right]}}
(\det [ e^{-2 \sigma}  \Lambda^{-2} Q^{\a}\,_\b]) 
\eea

We formally find that the path integral in both frames differ by an infinite factor which is a divergent power of $ e^{2 \sigma}$. However, this ignores the fact that we must regularise and renormalise the theory to obtain finite results. After this is done we will obtain a finite difference between the two path integrals.

\subsection{Regularisation and renormalisation}

In order for the expressions for the one-loop determinants to make sense we should introduce a UV cut off at the scale $\Lambda$ to regularise the functional integral and include counter terms.
The cutoff can be introduced using the Schwinger proper-time representation of the functional-trace \cite{Vassilevich:2003xt}
\bea
\Gamma = S_0 +  S_{\rm ct}(\Lambda) - \frac{1}{2} {\rm Tr }  \int_{1/\Lambda^2}^{\infty} ds  s^{-1} e^{- s\Delta} + {\rm Tr }  \int_{1/\Lambda^2}^{\infty} ds s^{-1} e^{- sQ}
\eea
where in the limit $\Lambda \to \infty$ the traces approach the unregulated form.
 The counter term $S_{\rm ct}(\Lambda)$ should be chosen such that $\Gamma$ is independent of the cutoff scale 
\bea
\frac{\partial \Gamma}{\partial \Lambda} = 0 \,.
\eea

In the $\tilde{\Phi}$ frame we can follow the same procedure and write
\bea
\tilde{\Gamma} = S_0 +  \tilde{S}_{\rm ct}(\Lambda)  - \frac{1}{2} {\rm Tr }  \int_{1/\Lambda^2}^{\infty} ds  s^{-1} e^{- s  e^{-2 \sigma}  \Delta} + {\rm Tr }  \int_{1/\Lambda^2}^{\infty} ds s^{-1} e^{- s e^{-2 \sigma} Q}
\eea
and again choose $\tilde{S}_{\rm ct}(\Lambda)$ such that $\tilde{\Gamma} $ is independent of $\Lambda$ 
\bea
\frac{\partial \tilde{\Gamma}}{\partial \Lambda} = 0\,.
\eea

An important observation to be made here is that by making the replacement $\Lambda \to \Lambda e^{\sigma}$ one can relate the difference between the effective action and the counter terms in the two frames by 
\bea \label{Lambda_replace}
\tilde{\Gamma}  -  \tilde{S}_{\rm ct}(\Lambda) =  \Gamma  -  S_{\rm ct}(\Lambda) |_{\Lambda \to \Lambda e^{ \sigma} } \,.
\eea

By a straight forward calculation it is then easy to show that the logarithmic dependence of the counter terms is given by\footnote{This approach to computing the one-loop effective action in curved space-time is commonly known as Schwinger-Dewitt technique \cite{DeWitt:1975ys,Barvinsky:1984jd} or Heat Kernel method \cite{Gilkey:1995mj,Vassilevich:2003xt}.}
 \bea
\Lambda \partial_\Lambda S_{\rm ct}(\Lambda)  = ...+ \frac{1}{(4 \pi )^2 } \int d^4x \sqrt{|g|}  ( B_{4}(\Delta) - 2 B_{4}(Q) )+ ...
 \eea
 and 
 \bea
\Lambda \partial_\Lambda \tilde{S}_{\rm ct} (\Lambda)    =  ... +\frac{1}{(4 \pi )^2 } \int d^4x \sqrt{|g|}  ( B_{4}( e^{-2 \sigma} \Delta) - 2 B_{4}( e^{-2 \sigma} Q) )+...
 \eea
 where the coefficients $B_{4}$, whose argument indicates the relevant differential operator, are the dimensionless heat kernel coefficient in four dimensions in the expansion
 \bea
{\rm Tr} \left(e^{-s D}\right) =   \frac{1}{(4 \pi s)^2 } \sum_{n} \int d^4x \sqrt{|g|} B_{2n}(D) s^n 
\eea 
for a given differential operator $D$.
However one can show \cite{Gilkey:1995mj,Vassilevich:2003xt} that although the operators differ in the two frames the coefficients agree 
 such that $ \sqrt{g}B_{4}( e^{-2 \sigma} \Delta) =  \sqrt{g} B_{4}( \Delta)$ and $ \sqrt{g}B_{4}( e^{-2 \sigma} Q) = \sqrt{g} B_{4}( Q)$.

 Thus the scheme independent renormalisation in both frames agree and one can identify the scheme independent counter terms in both frames
  \bea \label{counter_term}
 S_{\rm ct} =   \tilde{S}_{\rm ct}  =  ...  + \log(\Lambda/\mu)  \int d^4x \sqrt{|g|} B_{4} +  ...
 \eea
 where for brevity we define the sum of the heat kernel coefficients $B_4 \equiv B_{4}( \Delta) - 2B_{4}(Q)$.
 The ellipses in \eqref{counter_term} and previous formulas includes scheme dependent terms which have either a power law dependence on $\Lambda$ or vanish on-shell. Note that we have been forced to introduce a renormalization scale $\mu$ in order to cancel the divergence. This amounts to the fact that divergences will be the same in both frames, agreeing eventually with the results of \cite{Kamenshchik:2014waa}.

However, matching the counter terms in this way is not enough to conclude that the theories differ only by scheme dependent terms and are therefore physically equivalent. Instead we need to compare the renormalised effective actions, where finite terms might be relevant.
The relation \eqref{Lambda_replace} already indicates to that these finite terms will differ.
 In order to make the comparison, let us note that the regulated traces and the classical action $S_0$ are themselves independent of $\mu$. Consequently, there must be a physical scale 
 \bea
 M_{\rm phys} = M_{\rm phys}(\phi,g_{\mu\nu})\,,
 \eea
coming from the classical action and which may depend on the fields as well as the couplings, such that the logarithmic dependence on $\Lambda$ takes the form
\bea \label{Regulated_Trace_Log_M}
-\frac{1}{2} {\rm Tr }  \int_{1/\Lambda^2}^{\infty} ds  s^{-1} e^{- s  \Delta} + {\rm Tr }  \int_{1/\Lambda^2}^{\infty} ds s^{-1} e^{- s  Q}  =   ...  +    \int d^4x \sqrt{|g|}   \log (M_{\rm phys}/\Lambda) B_{4} + ... \,,
\eea
with $M_{\rm phys}$ compensating the dimension of $\Lambda$ in the argument of the logarithm.
In principle one should be able to calculate $M_{\rm phys}$.\footnote{The classical example of this is the Coleman-Weinberg potential \cite{1973PhRvD71888C}, where $M_{\rm phys}$ will be a combination of the mass and vacuum expectation value of the scalar field.}
After subtracting the counter term we will then have a finite contribution to the effective action given by  
\begin{align}
\Gamma \ni \int d^4x  \sqrt{|g|} \log   (M_{\rm phys}/\mu)   B_{4}.
\end{align}

Now if we consider the effective action in the $\tilde{\Phi}$ frame we see from \eqref{Lambda_replace} that $\Lambda$ is placed by $\Lambda e^{ \sigma} $ in \eqref{Regulated_Trace_Log_M} and thus 
\bea
\tilde{\Gamma} \ni    \int d^4x  \sqrt{|g|}  \log( e^{-\sigma} M_{\rm phys}/\mu) B_4 
\eea
which amounts to the replacement of the physical scale $M_{\rm phys}$ by
\bea \label{Mphys_relation}
 \tilde{M}_{\rm phys} =    e^{-\sigma} M_{\rm phys}.
\eea
One can then conclude that the finite effective actions in both frames differ by  
\bea \label{eq:relation_gammas}
  \tilde{\Gamma} - \Gamma = - \mathcal{A} + {\rm off\,shell\,terms} 
\eea
since after subtracting the counter terms there will remain a finite contribution
\bea \label{eq:frame_disc}
 \mathcal{A} =  \frac{1}{(4 \pi)^2}  \int d^4x \sqrt{|g|} \, \sigma(x)   B_4(x)\,,
\eea
which is present even after going on shell. We will refer to this quantity as {\it frame discriminant} hereinafter. 
An equivalent way to quantify the difference between quantising the theory in either frames follows from promoting $\mu$ to a field dependent scale via
\be \label{mu_transform}
\tilde{\Gamma} = \Gamma|_{\mu \to \bar{\mu} = e^{\sigma} \mu} \,.
\ee
As we shall explain in more detail in section~\ref{sec:discr_scalar_ten} this transformation resembles the transformation made in so called scale invariant renormalisation schemes.
However, we stress that \eqref{mu_transform} is much more general and applies to related theories quantised in the standard manner beginning in separate frames regardless of whether we have scale invariance.

Thus we can conclude that the two theories are inequivalent at the one-loop level and will therefore give different physical predictions, derived from the frame discriminant ${\cal A}$. This means that there is an ambiguity in the quantization of the theory related to the choice of the functional measure. This choice of functional measure can in turn be traced to a choice of which spacetime metric is declared to be the physical one. The frame discriminant $\mathcal{A}$ is finite and a function of the fields in the theory, so it will potentially generate new S-matrix transitions that were being disregarded before. Indeed, in the next section we will show how this piece solves the anomaly problem in the scale invariant scalar-tensor theory that we used as an example in Section \ref{sec:frame_anomaly}.

\section{The frame discriminant in scalar-tensor theories}\label{sec:discr_scalar_ten}
Now that we have presented a precise derivation of the frame discriminant, let us go back to the explicit example of a scalar-tensor theory introduced in section \ref{sec:frame_anomaly}, with actions in the Einstein and Jordan frames given by
\begin{align}
&S_E[\tilde{g}_{\m\n},\tilde{\phi}]=\int d^4 x \sqrt{|\tilde{g}|}\left(- M_p^2 \tilde{R} +\frac{1}{2}\partial_\mu \tilde{\phi}\partial^\mu\tilde{\phi} +\frac{\lambda}{4!}\frac{ M_p^4}{ \xi^2}\right)\\
&S_J[g_{\m\n},\phi]=\int d^4 x \sqrt{|g|}\left(-\xi \phi^2 R+\frac{1}{2}\partial_\mu \phi \partial^\m \phi +\frac{\lambda}{4!}\phi^4   \right)
\end{align}
where the variables are related by
\begin{align}\label{eq:transf2}
\tilde{g}_{\m\n}=\frac{\xi \phi^2}{M_p^2}g_{\m\n}, \qquad \tilde{\phi}=M_p\sqrt{\frac{1}{\xi}  + 12}\, \log\left(\frac{\phi}{m}\right)\,.
\end{align}

As we commented in section \ref{sec:frame_anomaly}, the standard quantization of this theory carried out in each frame leads to inequivalent theories due to the presence of an anomaly only in the Jordan frame. This is precisely a consequence of defining the functional measure in one of the frames, where we are thus choosing a preferred metric $\mathfrak{g}_{\mu\nu}$, as discussed in previous sections. If we want to rewrite the theory in any other frame, we need to transform this functional measure as well, picking up the finite contribution of the frame discriminant into the quantum effective action, which for this particular example will solve the clash with the scale anomaly, as we will see. Quantizing in any other frame without taking care of this represents, as previously discussed, a different choice of functional measure and thus \emph{a different quantum field theory.}

Here we can identify the issue by looking at the measures for both theories. Quantising in the Jordan frame, the line element of the field space metric is given by
\bea  \label{Jordan_meteric_in_Jordan_Frame}
C_{J ab} \delta \Phi^a \delta \Phi^b  = \Lambda^2 \int d^4x \sqrt{g} \left(  \frac{1}{4}  \xi \phi^2 g^{\mu\nu\alpha \beta}  \delta g_{\mu\nu} \delta g_{\alpha \beta}     -2 \xi  \phi g^{\mu\nu}   \delta g_{\mu\nu} \delta \phi   +  (1+ 4 \xi) \delta \phi \delta \phi  \right)
\eea
which is not scale invariant and hence we will have the usual scale anomaly. If we now were to make an innocuous change of variables we obtain 
\bea \label{Jordan_meteric_in_Einstein_Frame}
C_{J ab} \delta \Phi^a \delta \Phi^b   &=& C_{E ab} \delta \tilde{\Phi}^a \delta \tilde{\Phi}^b \nonumber \\
&=&   \Lambda^2 \int d^4x  \sqrt{ \tilde{g}}  \frac{ M_p^2}{ m^2 \xi}\,   {\rm exp}\left( -\frac{2\tilde{\phi}}{M_p \sqrt{\xi^{-1} +12}}\right)         \Bigg(     \frac{1}{4}   M_p^2     \tilde{g}^{\mu\nu\a \b}  \delta \tilde{g}_{\a\b}   \delta \tilde{g}_{\mu\nu}+     \delta \tilde{\phi} \delta \tilde{\phi}   \Bigg) 
\eea
which is the Jordan frame metric written Einstein frame variables. Notably \eqref{Jordan_meteric_in_Einstein_Frame} is not invariant under a shift of $\tilde{\phi}$: the scale anomaly in the Jordan frame has transmuted  into a shift anomaly in the Einstein frame as a consequence of us quantizing the theory in the Jordan frame.
Conversely if we quantize the theory in Einstein frame the field space metric is given by  
\bea
\tilde{C}_{E ab} \delta \tilde{\Phi}^a \delta \tilde{\Phi}^b  = \Lambda^2 \int d^4x \sqrt{\tilde{g}} \left(  \frac{1}{4}  M_p^2 \tilde{g}^{\mu\nu\alpha \beta}  \delta \tilde{g}_{\mu\nu} \delta \tilde{g}_{\alpha \beta}      +  \delta \tilde{\phi} \delta \tilde{\phi}  \right)
\eea
which is invariant under the shift symmetry for $\tilde{\phi}$  and thus we have no anomaly. 
Performing the change of variables, this time from the Einstein frame to the Jordan frame, we obtain the field space metric
\bea
\tilde{C}_{E ab} \delta \tilde{\Phi}^a \delta \tilde{\Phi}^b   &=& \tilde{C}_{J ab} \delta \Phi^a \delta \tilde\Phi^b \nonumber \\
&=&  \frac{\Lambda^2}{M_p^2} \int d^4x \sqrt{g} \phi^2 \xi \left(  \frac{1}{4}  \xi \phi^2 g^{\mu\nu\alpha \beta}  \delta g_{\mu\nu} \delta g_{\alpha \beta}     -2 \xi  \phi g^{\mu\nu}   \delta g_{\mu\nu} \delta \phi   +  (1+ 4 \xi) \delta \phi \delta \phi  \right)
\eea
which, in contrast to \eqref{Jordan_meteric_in_Jordan_Frame}, is scale invariant and hence we do not have a scale anomaly.
Thus we have the choice of two quantizations: the anomaly  free `Einstein frame quantization' and the anomalous  `Jordan frame quantization'. Picking the Einstein frame as the {\it preferred frame} in which to determine the measure will mean we remain anomaly free even if we ultimately use Jordan frame variables. 

Let us now work out the form of the frame discriminant to see how it preserves scale invariance of the level of the one-loop quantum effective action provided we pick the Einstein frame as the preferred frame.
The divergent part of the effective action in the Einstein frame can be computed at one-loop by the use of standard techniques. Here we show the results in dimensional regularization. We refrain to reproduce the details of such computation here and refer the reader to the literature instead, e.g. \cite{Barvinsky:1993zg,Steinwachs:2011zs,Alvarez:2014qca}. When evaluated on the mass-shell,
we have 

\begin{align} \label{Gamma_E}
\tilde{\Gamma}_E=-\frac{1}{(4\pi)^2}
\frac{71}{60}\int d^4x\sqrt{|\tilde{g}|}\,\tilde{C}_{\mu\nu\rho\sigma}\tilde{C}^{\mu\nu\rho\sigma} \log ( \mu /\tilde{M}_{\rm phys})\
\end{align}
where $C_{\mu\nu\rho\sigma}$ is the Weyl tensor of the manifold.
Here we are assuming a constant profile for the on-shell scalar field
\bea
\tilde{\phi} = {\rm const.}
\eea
and setting $\lambda = 0$
to simplify the discussion. In a more general case we would find similar results but the expressions would be longer and less transparent for our purposes here.
 Now since in the Einstein frame we have an unbroken shift symmetry we know that $\tilde{M}_{\rm phys}$ must be invariant under \eqref{eq:Shift} and thus for a constant $\tilde{\phi}$  the physical scale $\tilde{M}_{\rm phys}$ is independent of $\tilde{\phi}$ and only depends on the metric $\tilde{g}_{\mu\nu} = e^{2 \sigma(\phi)} g_{\mu\nu} $ where from \eqref{eq:transf2} $\sigma$ is given by 
\begin{align}
\sigma(\phi)=\frac{1}{2}\log\left(\frac{\xi \phi^2}{M_p^2}\right)=\log\left(\frac{\sqrt{\xi}\,\phi}{M_p}\right)\,.
\end{align}
Writing the action \eqref{Gamma_E} in the Jordan frame variables we then obtain\footnote{Here the subscript $J$ simply denotes which variables we are using while the tilde indicates that preferred frame is the Einstein frame}
\begin{align}
\nonumber \tilde{\Gamma}_J&=-\frac{1}{(4\pi)^2} \frac{71}{60}\int d^4x\sqrt{|g|}\,C_{\mu\nu\rho\sigma}C^{\mu\nu\rho\sigma}  \log (\mu/ \tilde{M}_{\rm phys}(  e^{2 \sigma} g_{\mu\nu})) 
\end{align}
which is scale invariant since the shift symmetry had simply transformed into the scale symmetry under the change of variables.
 We can equally write the effective action in Jordan variables   according to \eqref{eq:relation_gammas} as
\begin{align}
\tilde{\Gamma}_J=   \Gamma_J
 -{\cal A}
 \end{align}
in terms of the effective action 
\bea
\Gamma_J = -\frac{1}{(4\pi)^2}
\frac{71}{60}\int d^4x\sqrt{|g|}\,C_{\mu\nu\rho\sigma}C^{\mu\nu\rho\sigma} \log ( \mu /M_{\rm phys})
\eea
which we would obtain if we were to take Jordan frame as the preferred frame, and the frame discriminant   
\begin{align}
{\cal A}=\frac{1}{(4\pi)^2}\int d^4x \sqrt{|g|}\,\sigma B_4=\frac{1}{(4\pi)^2}\frac{71}{60}\int d^4x\sqrt{|g|}\, \log\left(\frac{\sqrt{\xi}\phi}{M_p}\right)C_{\mu\nu\rho\sigma}C^{\mu\nu\rho\sigma}\,.
\end{align}
In this form we see how the  frame discriminant comes to save frame equivalence and solves the problem with the scale anomaly.
First let us note that from the relation \eqref{Mphys_relation} and using that  $ \tilde{M}_{\rm phys}$ is independent of $\tilde{\phi}$ we have
\bea
\tilde{M}_{\rm phys}  =   \tilde{M}_{\rm phys}(\tilde{g}_{\mu\nu}) = e^{-\sigma(\phi)} M_{\rm phys} \implies   M_{\rm phys} =  e^{\sigma(\phi)}   \tilde{M}_{\rm phys}( e^{2 \sigma(\phi)} g_{\mu\nu}) 
\eea
Thus, unlike $\tilde{M}_{\rm phys}$,  under a scale transformation \eqref{eq:scale_inv}  the physical scale $M_{\rm phys}$  transforms non-trivially as
\bea
M_{\rm phys} \to  \Omega^{-1} M_{\rm phys}
\eea
If we now compute the conservation of the current for scale invariance we will find that $\Gamma_J$ induces what we called before the anomaly
\bea
\delta \Gamma_J &=& -  \delta \left(  \frac{1}{(4\pi)^2}
\frac{71}{60}\int d^4x\sqrt{|g|}\,C_{\mu\nu\rho\sigma}C^{\mu\nu\rho\sigma} \log ( \mu /M_{\rm phys}) \right) \nonumber \\   &=&-\frac{\omega}{(4\pi)^2}\frac{71}{60}\int d^4x\sqrt{|g|}\,C_{\mu\nu\rho\sigma}C^{\mu\nu\rho\sigma}
\eea
for a constant transformation with coefficient $\Omega=1+\omega+O(\omega^2)$. 
However by taking into account the discriminant,  which transforms precisely as
\bea
\delta \mathcal{A} &=& \delta \left( \frac{1}{(4\pi)^2}\frac{71}{60}\int d^4x\sqrt{|g|}\, \log\left(\frac{\sqrt{\xi}\phi}{M_p}\right)C_{\mu\nu\rho\sigma}C^{\mu\nu\rho\sigma} \right) \\
 &=& - \frac{\omega}{(4\pi)^2}\frac{71}{60}\int d^4x\sqrt{|g|}\,C_{\mu\nu\rho\sigma}C^{\mu\nu\rho\sigma} \,,
\eea
we find that now the total quantum effective action is invariant
\begin{align}
\delta\tilde{\Gamma}_J=0
\end{align}
and there is not anomalous current whatsoever!

What is happening here is that the S-matrix, and thus all physical properties, are defined by \emph{the frame in which we define the functional measure}, where we implicitly choose a preferred metric $\mathfrak{g}_{\mu\nu}$. In any other frame, the effective action must transform appropriately in order to preserve all physical statements and in particular all S-matrix amplitudes. Since there are no anomalously generated elements in the Einstein frame, our quantization process must preserve this condition in any other frame.

The role of the frame discriminant in this example is precisely to compensate the differences in the finite pieces of the quantum effective action between the two different frames, being those the origin of the scale anomaly. But this also means that the frame in which we choose to start is very important. If instead we were starting from the Jordan frame, where the anomaly is a physical effect, the frame discriminant would give us exactly the opposite effect to what we have shown here -- to generate the consequences of the scale anomaly in any other frame, in order to preserve all S-matrix elements. Of course, this effect is not restricted to theories with anomalous currents, but it appears whenever we do a non-linear redefinition of variables which affects the integration measure. In summary, a quantum field theory is not defined solely by the action, but also by the choice of integration measure or equivalently by the choice of preferred frame which selects the form of the measure.

\subsection{A comment on scale-invariant regularization}
Let us take a closer look to the expression for the quantum effective action $\tilde{\Gamma}_J$ in the Jordan frame where the preferred metric is the Einstein frame metric $\tilde{g}$. It is given by 

\begin{align}
\nonumber &\tilde{\Gamma}_J=-\frac{1}{(4\pi)^2}\frac{71}{60}\int d^4x\sqrt{|g|}\,\log(\mu/M_{\rm phys})C_{\mu\nu\rho\sigma}C^{\mu\nu\rho\sigma}-\frac{1}{(4\pi)^2}\frac{71}{60}\int d^4x\sqrt{|g|}\, \log\left(\frac{\sqrt{\xi}\phi}{M_p}\right)C_{\mu\nu\rho\sigma}C^{\mu\nu\rho\sigma}\\
&=-\frac{1}{(4\pi)^2}\frac{71}{60}\int d^4x\sqrt{|g|}\, \log\left(\frac{\sqrt{\xi}\mu\phi}{M_p M_{\rm phys}}\right)C_{\mu\nu\rho\sigma}C^{\mu\nu\rho\sigma}
\end{align}
Looking to the last expression, we can see that our result is identical to the standard renormalized effective action $\Gamma_J$ (the first term in the first line), when the Jordan frame metric $g_{\mu\nu}$ is the preferred one, if we define a new renormalization scale
\begin{align}
\bar{\mu}=  z \phi,\qquad z=\frac{\sqrt{\xi}\mu}{M_p}
\end{align}
so that
\begin{align}
\tilde{\Gamma}_J=-\frac{1}{(4\pi)^2}\frac{71}{60}\int d^4x\sqrt{|g|}\, \log\left(\bar{\mu}/M_{\rm phys}\right)C_{\mu\nu\rho\sigma}C^{\mu\nu\rho\sigma}
\end{align}
which is just a special case of the transformation \eqref{mu_transform}.

That is, if we introduce a renormalization scale which is \emph{field dependent}, with a parameter $z$ encoding the scheme independence\footnote{Indeed, scheme independence of this approach has been studied through the Callan-Symanzik equation in several works. See \cite{Ghilencea:2016ckm,Ghilencea:2015mza,Tamarit:2013vda}.} inherited from $\mu$. Moreover, once in the broken phase, which is the only phase in which both frames are even classically equivalent, we have $\phi=\langle\phi\rangle+\delta\phi$ and therefore
\begin{align}
\log (\bar{\mu})= \log(z \langle \phi \rangle)+\frac{\delta\phi}{ \langle \phi \rangle}-\frac{1}{2}\left(\frac{\delta\phi}{ \langle \phi \rangle}\right)^2+...
\end{align}

If we set $ \langle \phi \rangle=M_p/\sqrt{\xi}$ we recover the usual logarithmic term, which leads to the standard expression for the beta functions of the couplings in the quantum effective action, plus an infinite tail of non-renormalizable interactions. Incidentally, this precise value for the vacuum expectation value of the scalar field, which breaks spontaneously the scale symmetry, gives rise to an Einstein-Hilbert term in the action with the right Planck mass $M_p$.

This construction can be found in the literature under the name of \emph{scale-invariant regularization}, motivated by the search of a common solution to the hierarchy and cosmological constant problems altogether \cite{Shaposhnikov:2008xi,Shaposhnikov:2009nk,Shaposhnikov:2008xb,Bezrukov:2012hx,Percacci:2011uf,Codello:2012sn,Bars:2013yba,Ghilencea:2017yqv,Gorbunov:2013dqa,Armillis:2013wya,Gretsch:2013ooa,Tavares:2013dga,Wetterich:2019qzx}, as well as to the question of whether scale invariance can be preserved at the quantum level as a fundamental symmetry of Nature. Indeed, if one uses this regularization by substituting $\mu$ by $\bar{\mu}$ everywhere, scale invariance is preserved in the quantum effective action at all orders in the perturbative expansion. Then, both the hierarchy and cosmological constant problems seem to be solved at once thanks to the cancellation of radiative corrections to dimensionful quantities \cite{Wetterich:1987fm,Englert:1976ep}. Afterwards, the spontaneous breaking of the symmetry by $\langle \phi \rangle$ gives rise to the standard terms plus new interactions. Ways to trigger this spontaneous symmetry breaking from the point of view of cosmology have been also recently explored \cite{Ferreira:2018qss,Ferreira:2018itt,Ferreira:2016vsc}.  

Our arguments here seem to suggest that this regularization can be also understood as a consequence of choosing the Einstein frame as our preferred frame, thus forcing the scale anomaly to be absent to satisfy equivalence, thanks to the contributions of the frame discriminant. In the literature about frame equivalence and scale invariant regularization (see e.g \cite{Bezrukov:2010jz,Mooij:2018hew,Shaposhnikov:2018nnm} and references therein) this is normally described in terms of two different regularization prescriptions -- \emph{prescription I} refers to taking the renormalization scale $\mu$ to be constant in the Einstein frame and field-dependent in the Jordan frame, while \emph{prescription II} represents the opposite situation. This would correspond, in our language, to choose the preferred metric in the Einstein frame (\emph{prescription I}) or in the Jordan frame (\emph{prescription II}) in total agreement with previous results.

The fact that a scale invariant renormalization procedure corresponds to a non-standard quantisation with a scale invariant measure has been observed in \cite{Codello:2012sn} where the idea was to have a renomalisation scheme that preserves exact local scale invariance (i.e. Weyl invariance) by the introduction of a dilaton i.e. the field $\phi$. In this case one can view the dilaton as an auxiliary field and that the local scale invariance is `fake' since one can always gauge fix the dilaton to be a constant. 
 From the view point of frames gauge fixing the dilaton is tantamount to going to the Einstein frame where the shift symmetry is now local such that the action must be independent of  $\tilde{\phi}$ since the shift transformation is now $\tilde{\phi}(x) \to \tilde{\phi}(x) + C(x)$. 

\section{Discussion and conclusions}\label{sec:conclusions}
In this paper we have studied the problem of frame equivalence of a given Quantum Field Theory. While in classical physics it can be easily proven that stationary trajectories map to stationary trajectories under a non-singular change of variables (of frame), Quantum Field Theory requires the extra ingredient of defining the path integral measure. In the case of scalars, fermions and Yang-Mills fields, the integration measure is typically field independent\footnote{The exception is when the kinetic term in the action is non-canonical, for example in the case of a non-linear sigma model.}, but it is not the case anymore if we want to preserve diffeomorphism invariance when the metric is a dynamical degree of freedom. 
When we quantize a theory in the Einstein frame, where the matter is minimally coupled to gravity and the scalar has a canonical kinetic term, the measure will depend on the metric alone. However the measure obtained by quantizing a theory in the Jordan frame will depend on the scalar field in addition to the metric. What we have established in this paper is that, even after transforming the measure to take account of the Jacobian (a purely mathematical operation), the measures are not equivalent.
The frame where we choose to define the path integral matters, and defining the measure in different frames leads to \emph{different Quantum Field Theories}. Of course one could simply insist that the measures in both frames are equivalent, however this is only possible if the quantization in one of the frames would be non-standard. 

Once we decide the preferred frame where we define the integral measure, this will also establish \emph{any} physical conclusion of the theory. If for some reason we however want to describe it in a different set of variables, perhaps for symmetry or interpretation reasons, then we must carry on the effect of changing frames in the integration measure, together with transforming the action. However the resulting effective action will differ from the one which would result from choosing the second frame as the preferred frame to define the measure. 
We have shown that this difference can be evaluated in a way which is close to Fujikawa's method for the trace anomaly \cite{Fujikawa:1980vr} and that it reduces, in the case of conformal rescalings of the metric, to the need of adding a \emph{frame discriminant} contribution to the one-loop Quantum Effective Action in the transformed frame
\begin{align}
{\cal A}=\Gamma - \tilde{\Gamma} =\frac{1}{(4\pi)^2}\int d^4x\sqrt{|g|}\, \sigma(x) \mathcal{O}(x)
\end{align}
where $\sigma$ is the conformal factor driving the field redefinition and $ \mathcal{O}(x)$ contains the local counterterms of the theory given explicittly by the heat kernel coefficient  $B_4(x)$, which is easily computable by standard techniques.

Our findings are of specific  interest in the case of scalar-tensor theories of the general form\footnote{A particular theory of this kind of important relevance is Higgs Inflation \cite{Bezrukov:2007ep} where 
\begin{align}
F(R,\phi)=-\frac{M_p^2 + \xi \phi^2}{2}R, \qquad V(\phi)= -\frac{\lambda}{4}(h^2-v^2)
\end{align}
with $\lambda$ and $v$ being the self-coupling and vacuum expectation value of the Standard Model Higgs boson.}
\begin{align}
S_J=\int d^4x\sqrt{|g|}\, \left(\frac{1}{2}\partial_\mu \phi \partial^\mu \phi +F(R,\phi) +V(\phi)\right)
\end{align}

These models are often used to explain inflationary dynamics by taking them to the Einstein frame, where the gravitational fluctuations are driven by an Einstein-Hilbert term $-M_p^2 \tilde{R}$ and one can interpret the dynamics of the theory as that of a scalar field rolling down a potential. The field redefinition relating both frames will be generally non-linear, most likely including a conformal transformation similar to \eqref{eq:transf}, and will thus produce a non-trivial transformation of the integration measure, regardless of the symmetries of the action. In those cases, the quantum effective action will always pick up an extra finite piece needed to ensure equivalence, as given by our prescription.

For models in which the transformation is simply a conformal transformation $\tilde{g}_{\mu\nu}=e^{2\sigma}g_{\mu\nu}$, and if we assume that the \emph{preferred frame} where we define the integration measure is the Einstein frame, our results can be summarized in the fact that the local part of the one-loop renormalized quantum effective action will read in both frames
\begin{align}
&\tilde{\Gamma}_E=-\frac{1}{(4\pi)^2}\int d^4x\sqrt{\tilde{g}} \log(\mu / \tilde{M}_{\rm phys}) \tilde{{\cal O}}(\tilde{R}_{\mu\nu\alpha\beta},\tilde{\phi})\\
&\tilde{\Gamma}_J=-\frac{1}{(4\pi)^2}\int d^4x\sqrt{g} \log\left(\mu e^{\sigma} / M_{\rm phys}\right) {\cal O}(R_{\mu\nu\alpha\beta},\phi)
\end{align}
where the particular form of the counter-terms ${\cal O}(R_{\mu\nu\alpha\beta},\phi)$ will depend on the choice for $V(\phi)$ and $F(R,\phi)$ in the classical action. 
The factor of $e^{\sigma}$ in which multiplies $\mu$ in $\tilde{\Gamma}_J$ ensures that these are just the same effective action written in different variables and arises from properly transforming the path integral measure.
This schematic form will hold for any quantum field theory, regardless of its renormalizability \cite{Barvinsky:2017zlx}.

That is, when changing frames one should not only transform the divergences of the theory but also promote the renormalization scale $\mu$ to be \emph{field dependent} precisely by a conformal transformation\footnote{One can check that, provided that the metric transforms as $\tilde{g}_{\mu\nu}=e^{2\sigma}g_{\mu\nu}$ and after choosing a chart of coordinates, any energy scale of the theory must transform as $\tilde{E}=e^{-\sigma}E$ by dimensional analysis.}. This statement can be actually found in previous literature as a way to preserve the predictions of Higgs \cite{Bezrukov:2010jz} and Higgs-Dilaton \cite{Bezrukov:2012hx} inflation or under the name of \emph{scale invariant regularization}. Here we give an extra formal justification to this procedure from the request of frame equivalence of the path integral formulation.

We have shown, in particular, that the introduction of the frame discriminant for scalar-tensor theories solves the problem pointed out in \cite{Herrero-Valea:2016jzz} with the action \eqref{eq:jordan_frame}, whose naive quantization generates a scale anomaly in the Jordan frame which is absent in the Einstein frame. Inclusion of the frame discriminant precisely compensates this effect and enforces the effective action and all S-matrix elements in both frames to agree.

Our result here is however not restrained to scale-invariant theories, scalar-tensor theories or even to conformal transformations (although this is the most typical situation in literature) but it applies to any Quantum Field Theory where the metric is a dynamical degree of freedom and a change of frame is performed. This includes, among others, several models of inflation \cite{Bezrukov:2007ep,Bezrukov:2012hx,Hertzberg:2010dc,Buchdahl:1983zz,Starobinsky:1980te,Casas:2017wjh}, higher derivative \cite{Stelle:1977ry,Stelle:1976gc}, Lovelock \cite{Lovelock:1971yv} and $F(R)$ gravity \cite{DeFelice:2010aj}, the relation between the string frame and the Einstein frame\cite{Alvarez:2001qj,Dick:1998ke}, and the Weyl invariant formulations of Unimodular Gravity \cite{Alvarez:2015sba,Jirousek:2018ago}. If we want to extract dependable conclusions from the Quantum Effective Action on any of these theories, we must add the frame discriminant contribution whenever we perform a change of variables. Otherwise we might be missing important physical effects that could strongly modify our conclusions.

There are three main questions open for future research following the work in this paper. First, it would be useful to extend our arguments here beyond the one-loop approximation. In particular, it would be interesting  to understand if the relation between frame equivalence and scale invariant renormalization holds at all orders, providing thus a complete justification for the use of the latter. More broadly one should establish a consistent effective field theory incorporating the choice of the measure and incorporating all of its consequences. 

On the other hand, it is reasonable to ask if there is any physical argument to prefer one frame over another. Taking into account that the operators in the action and those generated by radiative corrections differ in different frames, one could think that the choice must be influenced by the UV completion of the models that we are studying. Indeed, if we had such completion at our disposal, the procedure to obtain a low energy effective field theory would be unique and it would single out a preferred expression for the action and variables to use. It would be thus interesting to understand if we can actually make a reasoning on the opposite direction. If we can use our results here to pinpoint a given action as preferred, this could give us information on the shape of the UV completion of our theory, which in particular might be relevant to understand new features of Quantum Gravity.

Finally, it would be useful to have an explicitly frame invariant effective action, following the spirit of the Unique Effective Action of Vilkovisky, including the frame discriminant as a built-in feature. This can be achieved by properly incorporating a frame invariant integration measure for the path integral into the definition of the effective action as we have outlined here.

\section*{Acknowledgements}
We are grateful to Fedor Bezrukov, Christopher T. Hill, Roberto Percacci,  Sergey Sibiryakov and Anna Tokareva for discussions and/or e-mail exchange. We also wish to thank Mikhail Shaposhnikov and Sander Mooij for useful comments on a previous version of this text. Our work has received support from the Tomalla Foundation and the Swiss National Science Foundation.
\appendix

\section{Gauge fixing}
\label{gauge_fixing_App}

Here we will prove the relations introduced in section \ref{sect4} by constructing the gauge fixing sector in the frames $\Phi$ and $\tilde{\Phi}$.

The actions in both frames are invariant under diffeomorphisms, which we can express as 
\bea
\Phi^a \to \Phi^a + K^a_{\a}[\Phi] \epsilon^{\a}
\eea
where the generator $K^a_{\a}$ is given by
\bea
K_\a^a = \begin{pmatrix}
 g_{\mu \nu,\alpha}   +  g_{\mu \alpha} \partial_\nu    + g_{\alpha \nu} \partial_\mu  \\
     \phi_{,\alpha}      \\
   \end{pmatrix}   \delta(x,y) \,.
\eea
where the comma denotes a partial derivative.
Now we consider expanding  $S[\Phi]$  and  $\tilde{S}[\tilde{\Phi}]$ around a solution to the equations of motion and adding background gauge fixing terms with
\bea
&&S_{\rm gf} =\frac{1}{2}  F^{\a}[\Phi] Y_{\a\b} F^{\b}[\Phi]\,,    \,\,\,\,\,\,\,\,\,\,\,\,\,   F^{\a}[\Phi] = F^{\a}_a \Phi^a\\
&&\tilde{S}_{\rm gf} =\frac{1}{2}  \tilde{F}^{\a}[\tilde{\Phi}] \tilde{Y}_{\a\b} \tilde{F}^{\b}[\tilde{\Phi}]\,,  
  \,\,\,\,\,\,\,\,\,\,\,\,\,   \tilde{F}^{\a}
[\tilde{\Phi}] = \tilde{F}^{\a}_{a} \tilde{\Phi}^{a}
\eea
Here $Y$ and $\tilde{Y}$ are needed to make the gauge fixing action covariant and we will choose them to be ultra-local and choose $ F^{\a}_a$ and $ \tilde{F}^{\a}_{a}$ to be first order derivatives operators. Since we introduce $Y$ and $\tilde{Y}$ the path integrals after gauge fixing have the form
\bea
\mathcal{Z} = \int d\omega \int d\bar{c} \int dc \int d \Phi \frac{\sqrt{\det C_{ab} }}{\sqrt{\det \eta_{\a\b}}}  e^{- S[\Phi] - \frac{1}{2} \int d^4x \omega^{\a} Y_{\a\b} \omega^{\b} } \delta (F^{\a} - \omega^{\a})    \sqrt{ \det Y_{\a \b}} e^{-  \int d^4x \bar{c}_{\a} Q^{\a}\,_\b c^{\b}}
 \eea
where we integrate $\omega$ and to obtain \eqref{Gauge_fixed_Z}. Let us note that the anti-ghost $\bar{c}_{\a}$ is a one-form density of weight one while the ghost $c^{\a}$ is a vector of weight zero  such that the  Fadeev-Popov operator  
\bea
Q^{\a}\,_\b = \frac{\delta F^{\a}}{\delta \Phi^a} K^{a}_{\b}
\eea 
is a Laplace-type operator 
\bea \label{Qform}
Q^{\a}\,_{\b} = - \delta^\a_\b \nabla^2   + \gamma^{\a\mu}_{\b} \nabla_{\mu} + w^{\a}\,_{\b} \,.
\eea
where $\delta^\a_\b = \delta^{\mu}\,_\nu \delta(x,y)$ is the identity.

The corresponding second order operator which drives quantum dynamics in the $\Phi$ frame will be
\bea
\mathcal{D}_{ab}[\Phi]   =   \frac{\delta^2 S[\Phi]}{\delta \Phi^a \delta \Phi^b} + F^{\a}_a Y_{\a\b} F^{\b}_b
\eea

It is convenient to choose the gauge fixing condition $\tilde{F}_{\a}$ such that we also have 
\bea \label{D_relation}
 \tilde{\mathcal{D}}_{ab}[\Phi]  =  \frac{\delta^2 S[\Phi]}{\delta \tilde{\Phi}^{a} \delta \tilde{\Phi}^{b}} + \tilde{F}^{\a}_a \tilde{Y}_{\a\b} \tilde{F}^{\b}_b
  = \frac{\delta \Phi^b}{\delta \tilde{\Phi}^{b}}    \frac{\delta  \Phi^a }{ \delta \tilde{\Phi}^{a}   }  \mathcal{D}_{ab}[\Phi] 
\eea
which implies that we have the relation
\bea \label{Sgf_relation}
\tilde{S}_{{\rm gf} ,ab} = \frac{\delta \Phi^b}{\delta \tilde{\Phi}^{b}}    \frac{\delta  \Phi^a }{ \delta \tilde{\Phi}^{a}   }  S_{{\rm gf} ,ab}  \,.
\eea
Additionally we wish to choose the minimal gauge such that the Hessian is of the form \eqref{Minimal_Hessian}. 
In the $\Phi$ frame the minimal gauge is achieved by choosing 
\bea
Y_{\mu \nu}(x,y) =  - U(\phi) \sqrt{\det g} g_{\mu\nu}  \delta(x,y)
\eea
and
\bea
F_a^\a = \begin{pmatrix}
   \left( g^{\mu  (\rho} \nabla^{\lambda)}  - \frac{1}{2} \nabla^{\mu} g^{\rho \lambda}\right)     \\
     -  \frac{U' }{U}   g^{\mu\nu} \nabla_{\nu}      \\
   \end{pmatrix}   \delta(x,y) \,.
\eea
Since we require \eqref{Sgf_relation} we can demand that 
\bea
\tilde{F}_{a}^{\beta} =  J^{\beta}\,_{\a} F_a^{\a}   \frac{\delta \Phi^a}{\delta \tilde{\Phi}^{a}   } ,   \,\,\,\,\,\,\,    \tilde{Y}_{\a\b}  = J^{-1} Y J^{-1}
\eea
where $J$ should be ultra-local.  The generators of diffeomorphisms $K^a_{\a}$ and $\tilde{K}^{a}_{\a}$ are vectors on the space of fields (this follows straight forwardly from there defintion) so we have
\bea
\tilde{K}^{a}_{\a} = \frac{\delta \tilde{\Phi}^{a}}{\delta \Phi^a} K^a_{\a}\,.
\eea 
We can then conclude that
\bea
\tilde{Q}^{\a}\,_\b =    J^{\a}\,_\gamma Q^{\gamma}\,_\b \,.
\eea
To fix $J^{\a}\,_\gamma$ we demand that $\tilde{Q}^{\a}\,_\b$ has the minimal form
\bea
\tilde{Q}^{\a}\,_{\b}  =  - \delta^\a_\b \tilde{\nabla}^2   + \tilde{\gamma}^{\a\mu}_{\b} \nabla_{\mu} + \tilde{w}^{\a}\,_{\b} 
\eea
for which it follows that   $J^{\a}\,_\b =  e^{-2 \sigma} \delta(x,y) \delta^\mu_\nu$ since $\tilde{g}^{\mu\nu} = e^{-2 \sigma} g^{\mu\nu}$ and thus we arrive at \eqref{FP_relation}.

\bibliography{frame_bib}{}
\bibliographystyle{unsrt}
\end{document}